\documentclass[11pt,a4paper]{article}
\usepackage{t1enc}
\usepackage[latin1]{inputenc}
\usepackage[english]{babel}
\usepackage{harvard}
\usepackage{amssymb,stmaryrd}

\newcommand{\N}{\mathbb N}

\newcommand{\beq}{\begin{equation}}
\newcommand{\eeq}{\end{equation}}
\newcommand{\beqarr}{\begin{eqnarray}}
\newcommand{\eeqarr}{\end{eqnarray}}
\newcommand{\beqa}{\begin{eqnarray*}}
\newcommand{\eeqa}{\end{eqnarray*}}
\begin{document}
\thispagestyle{empty}

\title{Higgs and  gravitational scalar fields together   induce Weyl gauge}
\author{Erhard Scholz\footnote{University of Wuppertal, Department C, Mathematics, and Interdisciplinary Centre for History and Philosophy of Science; \quad  scholz@math.uni-wuppertal.de}}
\date{July 24, 2014 } 
\maketitle
\begin{abstract}
A common biquadratic potential for the Higgs field $h$  and an additional scalar field $\phi$, non minimally coupled to gravity, is considered in locally scale symmetric approaches to standard model fields in curved spacetime. A common ground state of the two scalar fields exists and couples both fields to gravity, more precisely to  scalar curvature $R$. In  Einstein gauge ($\phi = const$, often called ``Einstein frame"), also $R$ is scaled to a constant. This condition makes perfect sense, even in the general case, in the Weyl geometric approach. There it has been called {\em Weyl gauge}, because it was first considered by Weyl in the different context of his original scale geometric theory of gravity of 1918. Now it seems to get new meaning as a combined effect of electroweak theory and gravity, and their common influence on atomic frequencies.
\end{abstract}

\subsubsection*{Introduction}
During the last few years, several authors have started to  study  the standard model of elementary particle physics (SM) in a form allowing point dependent scale transformations. Basically two closely related, although slightly different methods, are used, 
a  (purely) {\em conformal} approach \cite{Meissner/Nicolai,Bars/Steinhardt/Turok:2014} and a {\em Weyl geometric} one \cite{Nishino/Rajpoot:2004,Nishino/Rajpoot:2009,Scholz:Annalen,Quiros:2013,Quiros:2014}.
 Moreover, a structurally similar Lagrangian, invariant under rigid (global) scale symmetries is studied in \cite{Shapovnikov_ea:2009}. In all these works a scalar field $\phi$ of scale weight $w(\phi)=-1$,  non-minimally coupled to a modified Hilbert term of the Lagrangian,  $L_H = -\xi^2 |\phi|^2 R$, plays an important role  (density $\mathcal{L}_H= L_H \sqrt{|det g|}$).
 It  communicates with the scalar field $\Phi$ of the electroweak sector, the Higgs field, imported to ``curved'' space. 

The authors of the cited papers notice that in their respective environments the SM Lagrangian can easily be brought into  a (locally) scale invariant form. One only has to identify the  dimensional quadratic coefficient $\mu^2$ of the Higgs potential, up to a proportionality factor,  with the value of the gravitational scalar field squared, $\mu^2 \sim |\phi|^2$. Although  at first  sight that may seem surprising because a constant  is related to a field value, a second glance shows that there is no deeper problem related to  it. If one considers the  scale choice in which $\phi$ is scaled to a constant,   both perspectives can easily be brought into agreement.\footnote{In Shapovnikv/Zenh\"ausern's approach (global scale invariance) this requires a cosmological model with constant Riemannian scalar curvature, essentially DeSitter or anti-DeSitter space \cite{Shapovnikov_ea:2009}.}
 Then the coupling coefficient in $L_H$ can be chosen to coincide with Einstein gravity. This scale choice is often called  {\em Einstein gauge}.\footnote{In Jordan-Brans-Dicke theory one usually speaks of the ``Einstein frame''.}

 In both scale covariant approaches   the scalar field part of the Lagrangian $\mathcal{L}_{sc} =  L_{sc} \sqrt{|det\, {g}|}$ can be  written in similar form,  although with slightly different meanings of the terms\footnote{Signature  $sign(g_{\mu \nu}) = (1,3)\sim (+---)$.}
 \beq
 L_{sc} = - \frac{\xi^2 }{2}|\phi|^2 R + \frac{1}{2} D_{\nu}\phi^{\ast}D^{\nu}\phi +  \frac{1}{2} D_{\nu}\Phi^{\ast}D^{\nu}\Phi - V_{bi}(\phi, \Phi) \, , \label{L_sc}
 \eeq
where the biquadratic potential term is
 \beq
V_{bi}(\phi, \Phi) = \frac{\lambda}{4}(|\Phi|^2 - \alpha^2 \phi^2)^2 + \frac{\lambda'}{4} \phi^4 \, . \label{V_bi}
\eeq 
In the conformal approach $R$ denotes the Riemannian scalar curvature,  derivatives are taken with regard to the Levi-Civita connection (in any representative of the conformal metric), moreover $\xi^2 = -\frac{1}{6}$ for achieving conformal coupling with the kinetic term of $\phi$.\footnote{ Some authors include a direct coupling of the Higgs field to scalar curvature. Bars/Steinhardt/Turok generalize the Lagrangian, but even then retain the conformal coupling $\xi^2= \frac{1}{6}$.}
 In the Weyl geometric approach $R$ denotes {\em Weylian scalar curvature} which is scale covariant of weight $w(R)=-2$; the derivatives relate to the uniquely determined affine connection of the underlying Weyl structure and the coefficient $\xi^2$ underlies no constraints.
In the following, we shall work in the {\em Weyl geometric} approach and, in particular, with Weyl-Omote-Dirac gravity.\footnote{For Weyl geometry and Weyl structures, their derivatives  and  curvature expressions see, e.g., \cite{Gilkey_ea,Yuan/Huang:2013};  for Weyl-Omote-Dirac gravity  \cite{Omote:1971,Dirac:1973,Quiros:2014,Scholz:2014paving}, among many. A very provisional survey of Weyl geometry in physics after 1950 is given in  \cite{Scholz:MainzarXiv}. } 

The modified Hilbert term $L_H$ also contributes to the quadratic monomial of the potential of $\phi$ and has to be added to $V_{bi}$. As we are here dealing with  a classical consideration, we may  abbreviate the real norm of $\Phi$ (the square root of the expectation value of $<\Phi^{\ast},\Phi>$) by $h$,  $|\Phi|^2 = h^2 $. Thus the full potential of the two scalar fields is:
 \beq
 V(\phi, h)= \frac{1}{2}\xi^2 |\phi|^2 R +  \frac{\lambda}{4}(h^2 - \alpha^2 \phi^2)^2 + \frac{\lambda'}{4} \phi^4 \, \label{V(phi,h)} \,
 \eeq
 By obvious reasons we shall call $\phi$ the {\em gravitational scalar field}. The common biquadratic potential $V_{bi}$ couples the Higgs field to the gravitational scalar field and thus, indirectly,  to gravity.\footnote{Quiros and Bars/Steinhardt/Turok include a direct coupling of the Higgs field to the Hilbert term. The latter remark that it is ``ignorable (\ldots) since $H$ is tiny compared to the Planck scale'' \cite[6]{Bars/Steinhardt/Turok:2014}; similarly  Quiros. The authors of  \cite{Meissner/Nicolai} have different interests and do not consider the modified Hilbert term at all. }
 
 In part of the  mentioned literature, the gravitational component  in the scalar field potential is  given no particular attention \cite{Nishino/Rajpoot:2004,Meissner/Nicolai}, in another part it is investigated from the point of view  of inflationary models \cite{Nishino/Rajpoot:2009,Quiros:2013,Bars/Steinhardt/Turok:2014}, or  in a wider perspective \cite{Quiros:2014}. 
  We reconsider this point in the following and show that, under very natural assumptions on the coupling coefficients ($\xi^2 , \lambda, \lambda', \alpha $), the gravitational coupling of $\phi$ is crucial for the ground state of the gravitational scalar field  and, indirectly, for the ground state of the Higgs field (section 1).  This has important consequences for  measuring processes by atomic clocks  in the Weyl geometric approach to gravity. It gives new support for an assumption on scale gauge proposed by Weyl, in his discussion during the years 1918 -- 1920 with Einstein. He argued for a preferred scale gauge in which the Weyl geometric scalar cuvature is set to a constant  (section 2). It will be called  {\em Weyl gauge}.
 
   Weyl gauge will be of great relevance for our understanding of  cosmological models, if the Weyl geometric generalization of Einstein gravity is an adequate framwork for  placing the  SM fields into a  gravitional environment. The same can be said for more localized regions of spacetime in which Einstein gravity turns out to be   unreliable. 
   A short outlook on such questions is given in section 3.\footnote{The core of our argument  can be found already in \cite{Scholz:2014paving}. There it is embedded in a wider exposition of Weyl geometric gravity for a  readership interested in the philosophy of physics. Here we concentrate exclusively on the intertwinement of the gravitational scalar field with the Higgs field and the new light it sheds on the Weyl gauge.}  
 
 \subsubsection*{1. Common ground states of the two scalar fields }
At first we have to investigate, whether a common ground state of the two scalar fields exists, in other words, whether  there is  a (local) minimum of $V(\phi,h)$ in both variables. An easy calculation
 shows  that this is in fact the case. The function gradient $grad\, V = (\partial_{\phi}V, \partial_h V )$ vanishes for 
 \beq h_o = \alpha |\phi|_o, \quad \phi_o^2 =-\frac{\xi^2}{\lambda'}R\, ; \label{grad V = 0} \eeq
  moreover $V(\phi_o, h_o)$ is a local minimum.\footnote{$\partial_{\phi}V= R\xi^2 \phi- \alpha^2\lambda \phi (h^2- \alpha^2 \phi^2)+ \phi^3 \lambda', \; \partial_h V = \lambda h (h^2 - \alpha^2 \phi^2) $, and for $\phi_o, h_o$ as in eq, (\ref{grad V = 0})  $Hessian\,(V)_{|(\phi_o,h_o)} >0$ (positive definite).  }
  
  In our signature, the  scalar curvature $R$ of the most important cosmological models is  
negative, and the positivity condition for $\phi_o^2$ is ensured if  $\lambda'>0$.\footnote{Calculated in Riemannian geometry, the  Robertson-Walker models  with warp (expansion) function $f(\tau)$ and constant sectional curvature $\kappa$ of spatial folia have scalar curvature $_g \hspace{-0.2em}R= - 6\left( (\frac{f'}{f})^2 + \frac{f''}{f} + \frac{\kappa}{f^2}  \right)$ in signature $(1,3)\sim(+---)$. For $\kappa \geq 0$, or at best moderately negative sectional curvature, and accelerating or ``moderately contracting'' expansion, 
$_g \hspace{-0.2em}R < 0$. The import of such models into Weyl geometry adds scaling freedom and rescaling of the curvature by a positve, point dependent factor $\Omega^{-2}$ and does not change the sign. \label{fn curvature}} 
Under these conditions there are common ground states $\phi_o, h_o$ of the two fields in 
\beq \phi_o^2 =  -\frac{\xi^2}{\lambda'}R\,, \qquad  h_o^2 = \alpha^2 \phi_o^2 =  -\frac{\alpha^2 \xi^2}{\lambda'}R \,  \label{common ground state}
\eeq 
with $R<0,\, \lambda'>0$.

An important observation can be made immediately. In their ground states, the two scalar fields (squared) and Weyl geometric scalar curvature $R$ are strictly proportional to each other. A scale gauge in which one of the fields is scaled to a constant results in constant values for the other two. If the squared Higgs field, the expectation value  $<\Phi^{\ast},\Phi> = h^2 $, is scaled to a constant, we shall speak of the {\em Higgs gauge}. From eq. (\ref{common ground state}) we read off that all typical gauges introduced until here are equal:
\beq \mbox{\em Einstein gauge} =   \mbox{\em Weyl gauge} = \mbox{\em Higgs gauge}
\eeq  
For the constant values of the fields in Einstein-Weyl-Higgs gauge we shall use the notations $\phi_c, h_c, R_c$

Expecting an agreement with Einstein gravity (in the Riemannian limit) and with the electroweak energy scale, we derive the following constraints from the Lagrangian (\ref{L_sc}) and the common ground state (\ref{common ground state}):
\beq \xi^2 \phi_c^2 = M_{pl}^2 \approx (2.4 \cdot 10^{18}\, GeV)^2\, , \qquad \alpha^2 \phi_c^2 = v^2 \approx (246 \, Gev)^2\,  \label{constraints}
\eeq
Therefore
\beq \frac{\xi}{\alpha} \sim \frac{M_{pl}}{v} \sim 10^{16} \, ,  \quad \mbox{ the ew-Planck hierarchy factor.}
\eeq  
If, moreover, we assume $\lambda' \sim 1$,  (\ref{common ground state}) and (\ref{constraints})  show  that $\phi_c$ lies ``logarithmically in the middle''  between $R$ and $M_{pl}$, i.e., $\phi_c$ is the geometrical mean between the two:
\[ |R|^{\frac{1}{2}}  \quad  \stackrel{\xi/\sqrt{\lambda'}}{ \longrightarrow} \quad |\phi_c| \stackrel{\xi}{ \longrightarrow} \quad M_{Pl}
\] 
For many cosmological models (eventually in 'later' times) $R \sim 10\, H^2$, with Hubble parameter $H$ and $\hbar c\, H \sim 10^{-33}\, eV$.\footnote{For the estimates of $R$  see footnote \ref{fn curvature}.} For  $\lambda'\sim 1$, we find that the order of magnitude of the second hierachy factor (between the energy level of the scalar field's ground state and Planck energy) is $\xi \sim 10^{30}$.  Then the ew scale $v$ lies  near to the geometrical mean between $\phi_c$ and $M_{pl}$:
\[ [\hbar c]\, H  \quad  \stackrel{\xi}{ \longrightarrow} |\phi_c|  \quad  \stackrel{\alpha}{ \longrightarrow} \quad  v  \quad  \stackrel{\alpha'}{ \longrightarrow} \quad  M_{pl}\, ,
\] 
where $\xi = \alpha \alpha'$,  $\alpha \sim 10^{14}, \; \alpha' \sim 10^{16}$.

 The observational values of the Higgs mass $m_h\approx 126\, GeV$ and of $v\approx 256\, GeV$ constrain the effective (classical) value of $\lambda$ to $\lambda \approx 0.24$.\footnote{The tachyonic mass term of the Higgs field $\frac{\lambda}{4}\alpha^2 |\phi_c|^2 =\frac{\lambda}{2}v^2$ turns into a real mass term for the Higgs excitation, $m_h^2= \lambda v^2$, thus the value vor $\lambda$.} 
At first glance one might expect that $\lambda'$ could be similarly constrained by  dark energy considerations. But this is not the case, as one can  check by inspecting the changes in the energy tensor  of the scalar sector after introducing the Higgs field.\footnote{\cite[sect. 4.6, 5.3 ]{Scholz:2014paving}} 
The often discussed question why the quartic term of the Higgs field does not dominate gravitational {\em vacuum energy} in the cosmological term finds  a  convincing { explanation on the classical level}: In the common ground state of the scalar fields the first term on the r.h.s of  eq. (\ref{V_bi}) vanishes and only the purely quartic one in $\phi$ survives.\footnote{Moreover, while in Einstein gravity 
the  cosmological constant term results in the anomalous feature of vacuum energy of being able to influence the dynamics of matter and geometry without back-racting to them, this problematic feature is dissolved here (like in other Jordan-Brans-Dicke like approaches). More details in \cite[sec. 5.3]{Scholz:2014paving}.}
 
The fermion masses of the SM become now   $m_f = \mu_f v = \mu_f |\Phi_c|=\mu_f \alpha \phi_c $, with  $\mu_f$ the fermion  Yukawa coupling coefficient. In the light of scaling they acquire, in a natural way, a  scale covariant expression of weight $w(m_f)    =-1$
\[ m_f = \mu_f |\Phi_o|=  \mu_f \alpha \phi_o \, . \]
 
 \subsubsection*{2. Weyl gauge reconsidered }
The proportionality between the (squared) scalar field  with the Weyl geometric scalar curvature has most important consequences for the understanding of measurement processes.
Quantum mechanics  describes how atomic  spectra depend on the mass of the electron.
For example, the energy eigenvalues of the Balmer series in the hydrogen atom are governed by the Rydberg constant $R_{ryd}$,
\beq E_n =  - R_{ryd} \frac{1}{n^2} \, , \qquad  n \in \N \, . \label{Balmer}\eeq
The latter (expressed in electrostatic units) depends on  the fine structure constant $\alpha_f $ and on the electron mass,   thus finally on the norm of Higgs field:\footnote{Vacuum permissivity $\epsilon _o = (4\pi)^{-1}$; then $e^2=2 \alpha_f \epsilon _o h c = \alpha_f \hbar c$. }
\beq R_{ryd} = \frac{e^4 \, m_e}{2 \hbar^2}= \frac{\alpha_f ^2}{2} m_e c^2 = \frac{\alpha_f ^2}{2}  {\mu _e} \, v c^2\,  \label{Rydberg constant I}
\eeq
Although this equation is a classical idealization which has to be scrutinized after  field quantization,  the  masses  of elementary fermions  depend, in our scale covariant approach, on  indirect coupling to gravity. The Rydberg ``constant'' turns into a scale covariant quantity of weight $-1$ and  scales with $\phi$, while the electron charge is considered as a ``true'' (nonscaling) constant.  In terms of (\ref{common ground state}) it is 
 \beq R_{ryd} = \frac{\alpha_f ^2}{2}  \, \mu _e \, h_o c^2 = \frac{\alpha_f ^2}{2} \alpha \, \mu _e \, \phi_o\, c^2\, .  \label{Rydberg constant II}
 \eeq 
 In Einstein-Weyl-Higgs gauge  the Rydberg factor is scaled to a  constant.
Similarly, the usual atomic unit of length for a nucleus of charge number $Z$,  the Bohr radius $l_{Bohr} = \frac{\hbar}{Ze^2 m_e} $, gets  rescaled just as well like $\phi^{-1}$ and is point independent in Einstein-Weyl-Higgs gauge.

In this way, typical {\em atomic time intervals} (``clocks'') and {\em atomic distances} (``rods'') are {\em regulated by the ew scalar field's ground state} $|h_o|$, thus by $\phi_o$  and finally by the Weylian scalar curvature (eq. (\ref{common ground state})).  Under the assumptions of section 2, 
a definition of units for central physical magnitudes like in the new SI rules establishes a measurement system in which {\em  the values of $|h|$  and $\phi$ are set to  constants by convention}.\footnote{The present  revision of the  international standard system SI is heading toward implementing  measurement definitions with time as the only fundamental unit, $u_T = 1\, s$ such that ``the ground state hyperfine splitting frequency of the caesium 133 atom  $\Delta\nu(^{133}Cs )_{\mbox{hfs}}$ is exactly $9\,192\,631\,770$ hertz''  \cite[24f.]{SI:2011}. In the ``New SI'', four of the SI base units, namely the kilogram, the ampere, the kelvin and the mole, will be redefined in terms of invariants of nature; the new definitions will be based on fixed numerical values of the Planck constant, the elementary charge, the Boltzmann constant, and the Avogadro constant (www.bipm.org/en/si/new$_{-}$si/). The redefinition of the meter in terms of the basic time unit by means of the fundamental constant $c$ was implemented already in 1983. Point dependence of the time unit because of locally varying gravitational potential will be inbuilt in this system. For practical purposes it can be outlevelled by reference to the {\em SI  second on the geoid} (standardized by the International Earth Rotation and Reference Systems Service IERS).  \label{fn SI} }

The scaling condition of the Einstein-Weyl-Higgs gauge  and (\ref{Rydberg constant II})  shed surprising new light on an ad hoc assumption introduced by Weyl during his 1918 discussion with Einstein. Weyl conjectured that  atomic spectra, and with them rods and  clocks,  adjust  to the ``radius of the curvature of the world'' \cite[309]{Weyl:STM}. In his view, natural length units are  chosen in such a way that scalar curvature is scaled to a constant,  the defining condition of  what we  call {\em Weyl gauge}. 
In the fourth edition of {\em Raum - Zeit - Materie} (translated into English by H.L. Brose) he wrote:
\begin{quote}
In the same way, obviously, the length of a measuring rod is determined by adjustment; for it would be impossible to give to {\bf this} rod at {\bf this} point of the field any length, say two or three times as great as the one that it now has, in the way that I can prescribe its direction arbitrarily. The world-curvature makes it theoretically possible to determine a length by adjustment. In consequence of this constitution the rod assumes a length which has such and such a value in relation to the radius of curvature of the world. \cite[308f.]{Weyl:STM}
\end{quote}

The electroweak link  explored here  underpins a feature of Weyl geometric gravity, which was introduced  Weyl  in a kind of ``a priori'' speculative move. About a  year after the time of the last quote, in the 5th (German) edition of {\em Raum - Zeit - Materie}, Weyl already called upon   Bohr's  atom model as a first step  towards  clarifying  his scaling conjecture:
\begin{quote}
Bohr's theory of the atom shows that the radii of the circular orbits of the electrons in the atom and the frequencies of the emitted light are determined by the constitution of the atom, by charge and mass of electron and the atomic nucleus, and Planck's action quantum.\footnote{``Die Bohrsche Atomtheorie zeigt, da\ss{} die Radien der Kreisbahnen, welche die Elektronen im Atom beschreiben und die Frequenzen des ausgesendeten Lichts sich unter Ber\"ucksichtigung der Konstitution des Atoms bestimmen aus dem Planckschen Wirkungsquantum, aus Ladung und Masse von Elektron und Atomkern \ldots'' \cite[298]{Weyl:RZM5}.}
\end{quote}
At the time when this was written, Bohr had already derived (\ref{Balmer})  and (\ref{Rydberg constant I}) for the Balmer series of the hydrogen atom and for the Rydberg constant \cite[201]{Pais:Inward}. Weyl saw, at first,   no reason to give up his scale gauge geometry. He  rather continued:
\begin{quote}
The most recent development in atomic physics has made it likely that the electron and the hydrogen nucleus are the fundamental constituents of all matter; all electrons have the same charge and mass, and the same is true for all hydrogen nuclei. From this it follows with all evidence that {\em the masses of atoms, periods of clocks and lengths of measuring rods are not preserved by some tendency of persistence; it rather is a result of some equilibrium state determined by the constitution of the structure (Gebilde), onto which it adjusts so to speak at every moment anew} (emphasis in original).\footnote{``Die neueste Entwicklung der Atomphysik hat es wahrscheinlich
gemacht, da\ss{}  die Urbestandteile aller Materie das Elektron und
der Wasserstoffkern sind; alle Elektronen haben die gleiche Ladung und
Masse, ebenso alle Wasserstoffkerne. Daraus geht mit aller Evidenz hervor,
da\ss{} {\em sich die Atommassen, Uhrperioden und Ma\ss{}stabl\"angen nicht durch
irgendeine Beharrungstendenz erhalten; sondern es handelt sich da um
einen durch die Konstitution des Gebildes bestimmten Gleichgewichtszustand,
auf den es sich sozusagen in jedem Augenblick neu einstellt}.'' (loc. cit., emph. in or.,  298)}
\end{quote}
The claim that ``it follows with all evidence'' was, of course, an overstatement. It is well known how  Weyl himself shifted his gauge concept from scale to phase only a few years later (in the years 1928-1929). After this shift he reinterpreted the Bohr frequency condition. In later discussions he referred to it as an argument {\em against} the physicality of his scale gauge idea.\footnote{Compare, for example, Weyl's  remarks in \cite[83]{Weyl:PMNEnglish}.}

This shift gives evidence to a paradoxical double face of Weyl's remarks with regard to the Bohr frequency condition.  For Weyl it may have contained a germ for the later distantiation from his first gauge theory, hidden behind an all too strong rhetoric of ``evidence''. But now  it appears again in a completely new light. 
Read in a systematical   perspective, Weyl's remarks from 1922/23 can  now even appear as foreshadowing  {\em  a  halfway marker  on the road towards  a  bridge between gravity and atomic physics}. Whether this bridge resists  depends, of course, on the  answer to the question whether or  not the link discussed here  between the scalar fields of gravity and  of ew theory  is  realistic (``physical''). This question is open for further research. 

At the end of the 1920s there was no chance for anticipating the electroweak pillar of the bridge. Historically, Weyl was completely right in considering the Bohr frequency condition as an indicator that his early scale gauge geometry could not be upheld  as a physical theory in its original form.
 Weyl's original interpretation of the scale connection as the electromagnetic potential  became obsolete  in the  1920s, but 
 his ad hoc hypothesis that {\em Weyl gauge} indicates measurements by material clocks and ``rods''  most directly may now get new support.

\subsubsection*{3. Discussion and outlook  }
In the present approach the two intertwined scalar fields $h$ and $\phi$ adapt in their ground states to the Weyl geometric scalar curvature. This means that there is {\em no complete  decoupling} of the electroweak sector from gravity at low energy, as often stated. 
The  {\em two scalar fields}  combine gravitationally like {\em twins}. Taken together they induce a kind of ``spontaneous breaking'' of (local) scale symmetry. Some of these aspects have already been discussed, in slightly different form  in \cite{Nishino/Rajpoot:2009,Quiros:2013,Quiros:2014} and similarly,   though obliquely if one considers it from  the Weyl geometric viewpoint,  in \cite{Bars/Steinhardt/Turok:2014} or \cite{Shapovnikov_ea:2009}. 

Here we have concentrated  on the role of Weylian scalar curvature. We found that the latter regulates the ground states of the twin scalar fields through the modified Hilbert term, which gives  a gravitational contribution to the potential of the scalar fields. If this term is effective for the expectation value of the Higgs field, measurements by atomic or other clocks indicate a Weyl gauged metric, rather than a Riemann gauged one.\footnote{In a Weyl geometry with vanishing scale curvature there is a scale gauge in which the scale connection (``Weyl field'') vanishes. Then the Weyl metric looks essentially like a Riemannian one. This gauge is called {\em Riemann gauge}. For the interplay of scale gauge indendence of observable quantities and Einstein-Weyl gauge  as most direct expression of them see \cite[sec. 2.3]{Scholz:Annalen}.} 
That has important consequences for cosmology and strong gravity regions of spacetime. Some of them are indicated in \cite{Quiros:2014}. But even stronger than considered there, Weyl gauge as the most direct expression of measured quantities would put a huge question mark behind models with accelerated or even ``inflationary'' expansion, because the latter implies high scalar curvatur in Riemann gauge.\footnote{See fn \ref{fn curvature}.}
Weyl gauge could even put ``expansion'' as such  in question, at least as a matter of fact, although it would continue to be illustrative  for the study of Riemann gauged cosmological models.\footnote{\cite{Wetterich:2013,Scholz:FoP,Scholz:2014paving}.}
 
 At the end, we have to touch the problem of quantization. It is not yet completely clear which features of local scale symmetry survive quantization. The widely spread view that the trace anomaly breaks local scale symmetry on the quantum level anyhow, resides on the misunderstanding that local scaling is the same as localized dilatations of Minkowski space.  But local rescalings operate on the field spaces only, not on the spacetime degrees of freedom. Thus the reason for the trace anomaly does not translate to local scale invariance \cite[2]{Bars/Steinhardt/Turok:2014}. Other quantum anomalies do arise, but already in the 1970s some experts have argued  that these can be cancelled by appropriate counter terms, at least up to arbitrary finite order \cite{Englert_ea:1976}. In more recent work based on functional integral methods it has been   shown that, with a suitable quantization procedure,\footnote{Method of point-dependent cut-off proportional to the gravitational scalar field.} 
 local scale invariance can be maintained on the quantum level and holds along the whole renormalization flow. This has been shown for an external metric, and a dynamical one with conformal coupling to the scalar field \cite{Percacci:RG_flow,Codello_ea:2013}. The authors of the second work conclude that they ``have shown in complete generality that one can define a flow of Weyl-invariant actions whose IR endpoint is a Weyl-invariant effective action'' \cite[20]{Codello_ea:2013}.
 
Many questions are still  open: Does the result of \cite{Codello_ea:2013} translate to the general case of Weyl geometric coupling of the scalar field? Can scale invariance be implemented into other approaches to quantization (in particular the mathematically more controlled ones)? What are the consequences of local scale symmetry on the quantum level for renormalizability?
What about unitarity, etc.?
 More investigations are needed. 
 
 But  there is no  good reason to discard the Weyl geometric approach to gravity because, as has often been  claimed in verbal communications, scale invariance is ``broken during quantization anyhow''. It could be just the other way round. Scale invariant quantization procedures may turn out helpful for a better understanding of how gravity and the standard model fields can go together.  

\small

 \bibliographystyle{apsr}
 \bibliography{a_lit_hist,a_lit_mathsci}

\end{document}